\documentclass[aps,prl,twocolumn]{revtex4-1}  

\usepackage{amsmath}    
\usepackage{amssymb}
\usepackage{graphicx}   
\usepackage{color}
\usepackage{subfigure}
\usepackage{cases}
\usepackage{epstopdf}

\begin{document}

\title{Anomalous exponents at the onset of an instability}

\author{  F. P\'etr\'elis, A. Alexakis}
\affiliation{ Laboratoire de Physique Statistique, \'Ecole Normale Sup\'erieure, CNRS, Universit\'e P. et M. Curie, Universit\'e Paris Diderot, 24 rue Lhomond, F-75005 Paris (France)}


\date{\today}

\begin{abstract}

Critical exponents are calculated {\it exactly} at the onset of an instability, using  {\it asymptotic expansion}-techniques.  
When the unstable mode is subject to multiplicative noise whose spectrum at zero 
frequency vanishes,  we show that the critical behavior  can be anomalous, {\it i.e.} the mode amplitude $X$ scales with departure from onset $\mu$ as $\langle X \rangle  \propto \mu^{\beta}$ with an exponent $\beta$ different from its deterministic value.  This behavior is observed in a direct numerical simulation of the dynamo instability and our results provide a possible  explanation to recent experimental observations.  

\end{abstract}


\maketitle


In the vicinity of a continuous phase transition,  the amplitude of the order parameter,
 say $M$, increases with the departure from the critical point, say  $\epsilon$,  as a power law, {\it i.e.}
 $M\propto \epsilon^\beta$. Mean-field theories predict simple rational numbers for the exponent $\beta$ (for instance $1/2$ for systems with cubic nonlinearities). 
It has been realized for a long time  that, because of  thermal fluctuations,  the power law may differ from this mean-field prediction \cite{Kada}. The exponents are then said to be anomalous.  Using renormalization-group techniques,  their value  can be calculated as a perturbative expansion in the critical dimension minus the spatial dimension of the system \cite{Wilson}.

Similarly, in the vicinity of a continuous instability in an out of equilibrium system, the amplitude of the unstable mode, say $X$, grows with the departure from onset, say $\mu$, 
as a power law  $\langle X \rangle \propto \mu^{\beta}$ 
(where the angular brackets denote  time-average).  Dynamical systems obtained using  normal form theory \cite{GH} provide simple rational values for $\beta$ (usually $1/2$ when the problem has the $X\rightarrow -X$ symmetry, $1/4$ at the tricritical point where the cubic nonlinearity vanishes and so on). Guided by the  phase transition observations, one may expect that fluctuations  shift the  exponent $\beta$  away from its mean-field value. 
Somehow surprisingly, the overwhelming majority of experiments on instabilities reports simple rational values in agreement with the mean-field prediction for $\beta$: anomalous exponents seem  not to be measured  in this context  \cite{kazumasa,Oh}.  
In a recent  experiment in a turbulent flow of liquid sodium, the dynamo instability has
been observed and some measurements  indicate that the first moment  of the magnetic field displays an exponent $0.77$ \cite{VKS}. It is possible that experimental biases are responsible for this observation: the instability is slightly imperfect and the numerical value of the exponent is then highly sensitive to the accuracy of determination of the onset. Another appealing possible explanation is that the turbulent fluctuations of the flow lead to the anomalous exponent \cite{GAFD}. With the latter in mind, we now describe a canonical model that leads to anomalous behavior similar to the one measured in the dynamo instability.

\begin{figure*}[t!]
\begin{center}
\centerline{
\includegraphics[width=6.0cm,height=2.7cm]{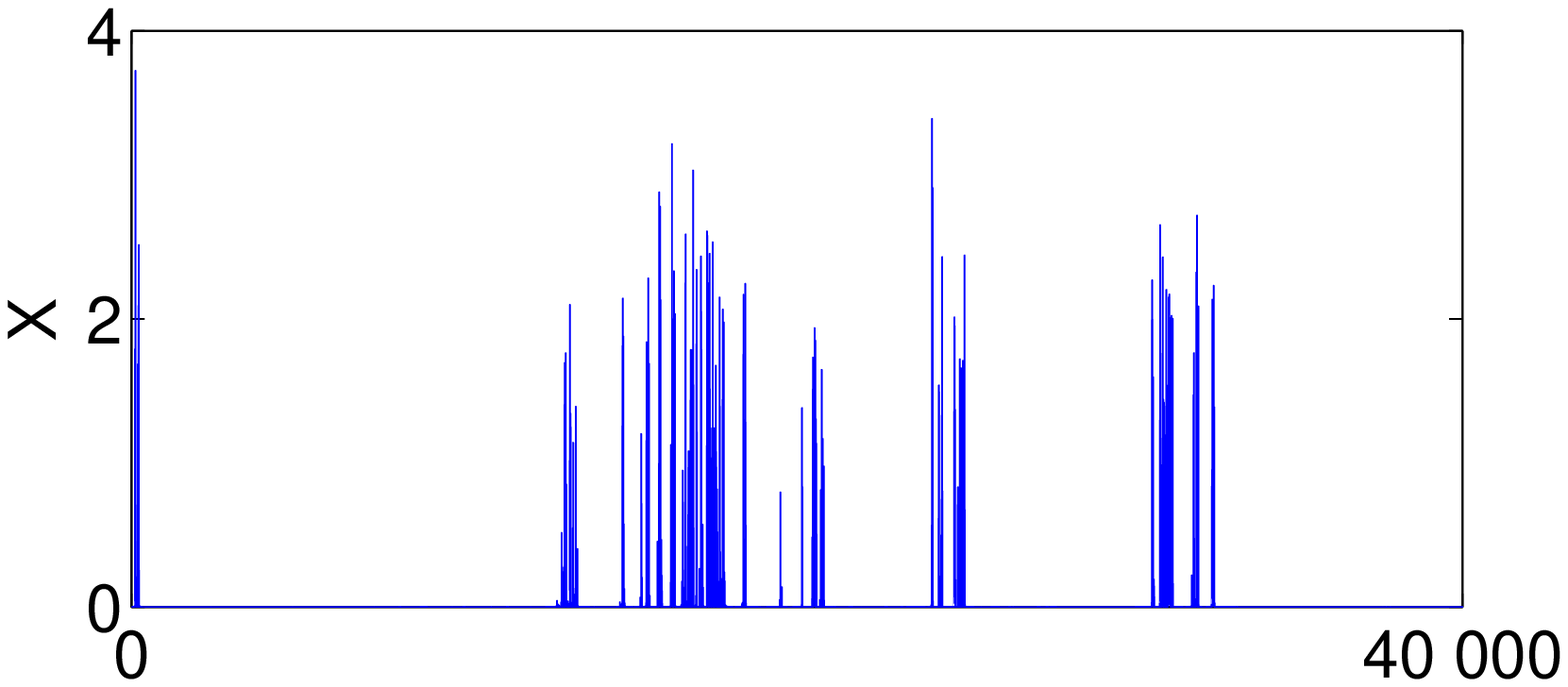}    
\includegraphics[width=6.0cm,height=2.7cm]{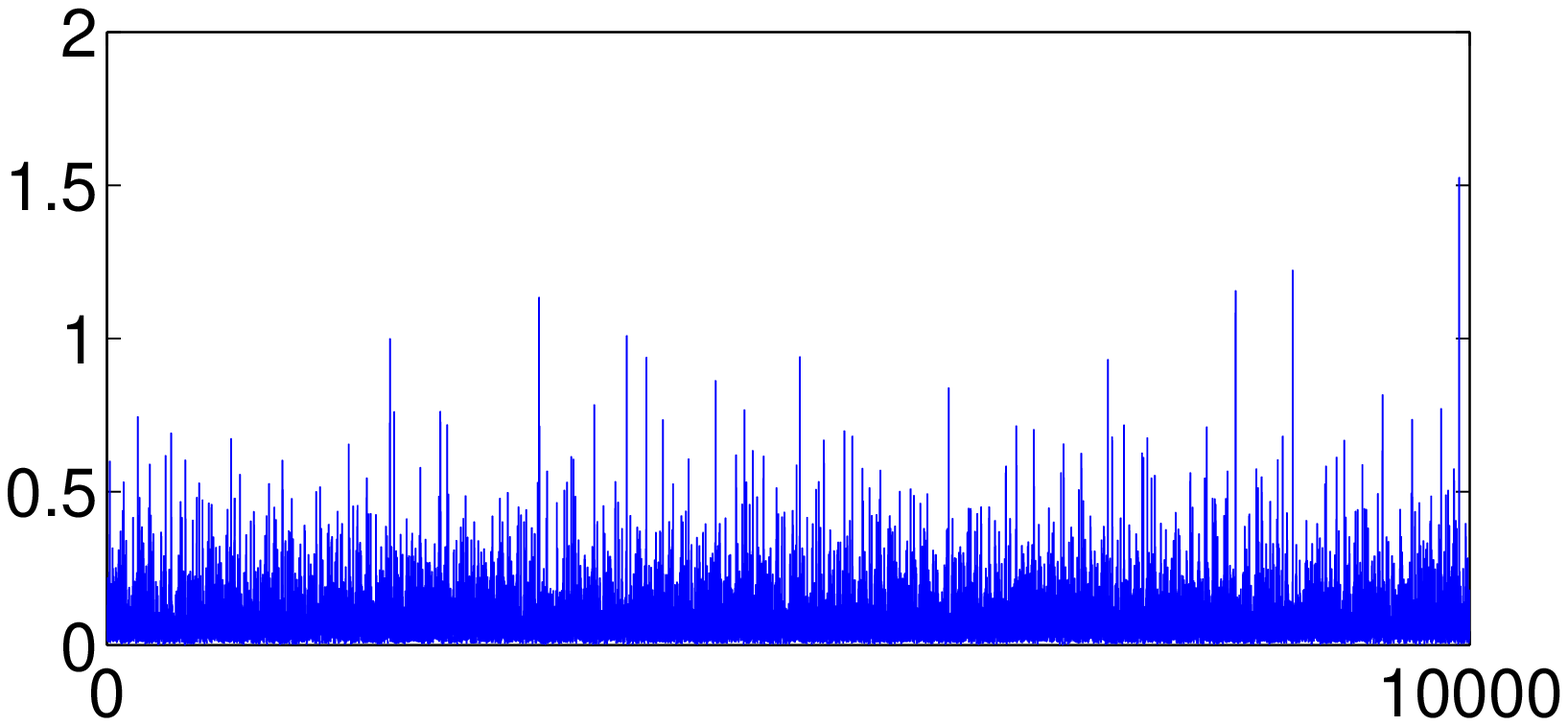}    
\includegraphics[width=6.0cm,height=2.7cm]{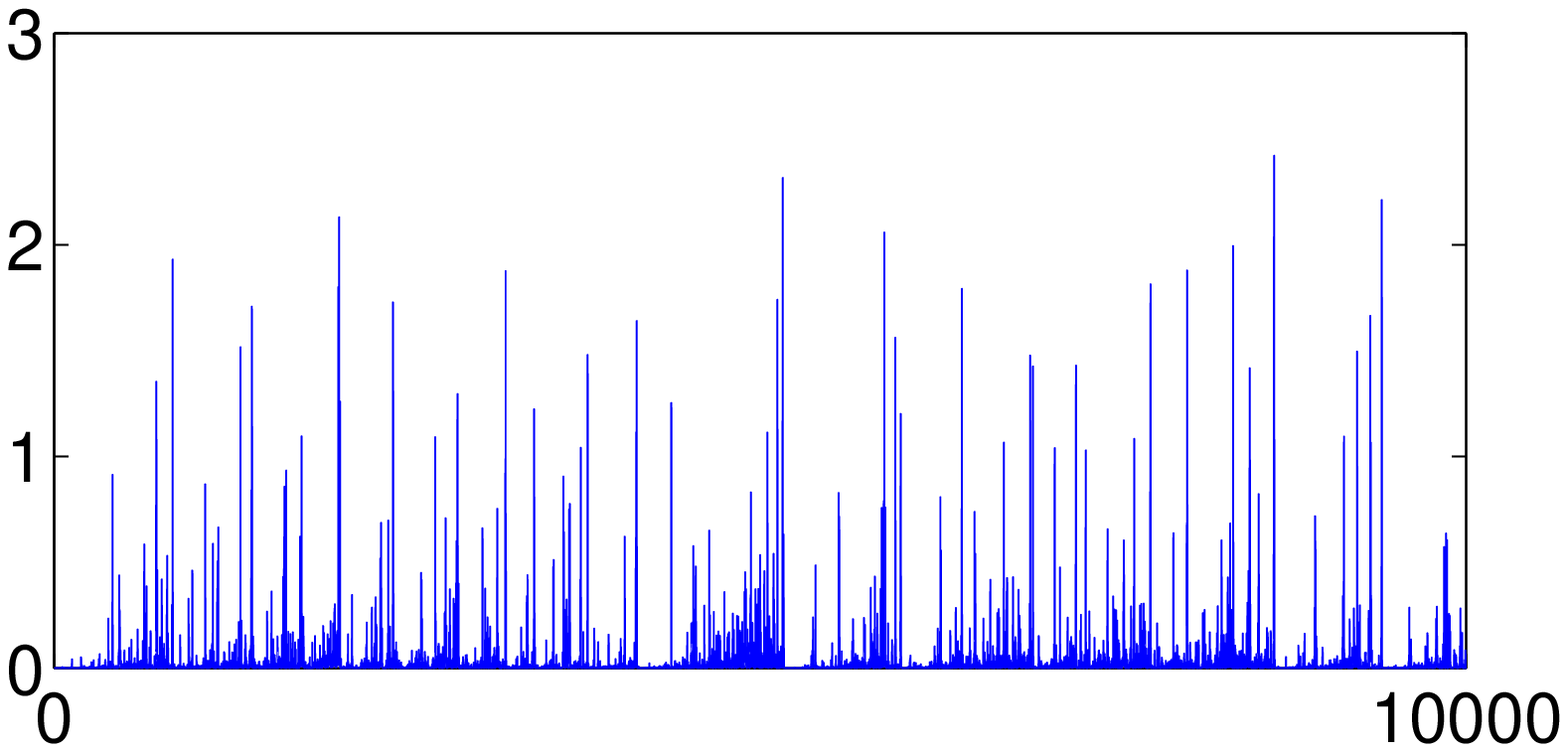}         }
\centerline{
\includegraphics[width=6.0cm,height=2.7cm]{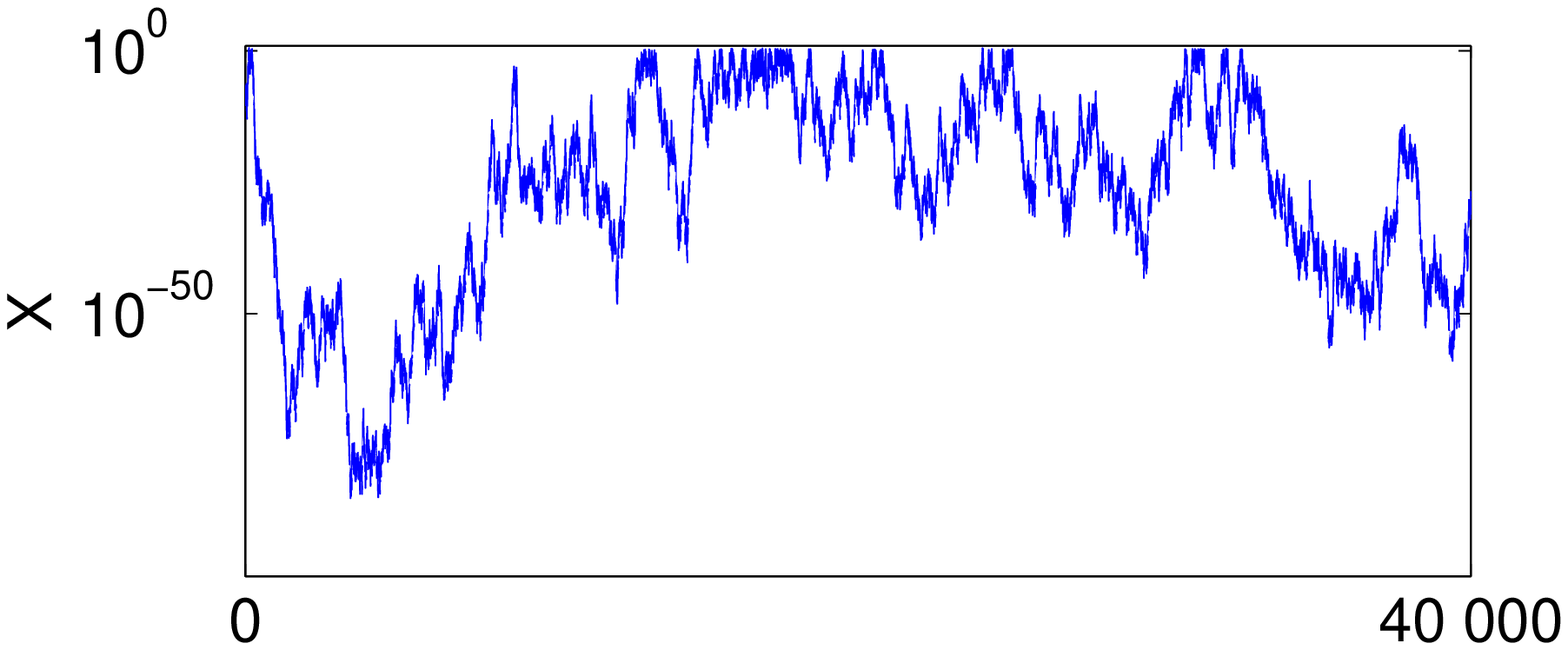} 
\includegraphics[width=6.0cm,height=2.7cm]{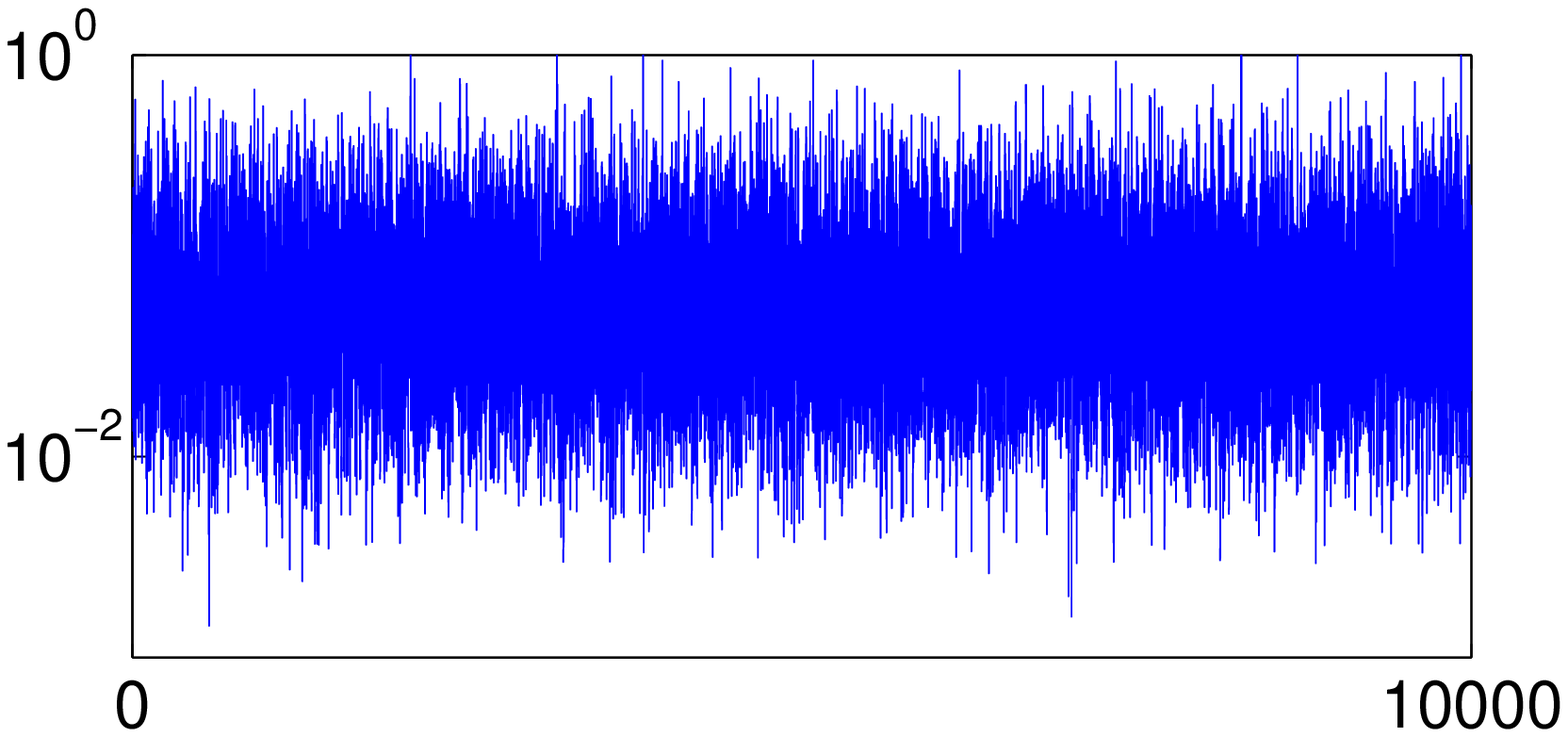} 
\includegraphics[width=6.0cm,height=2.7cm]{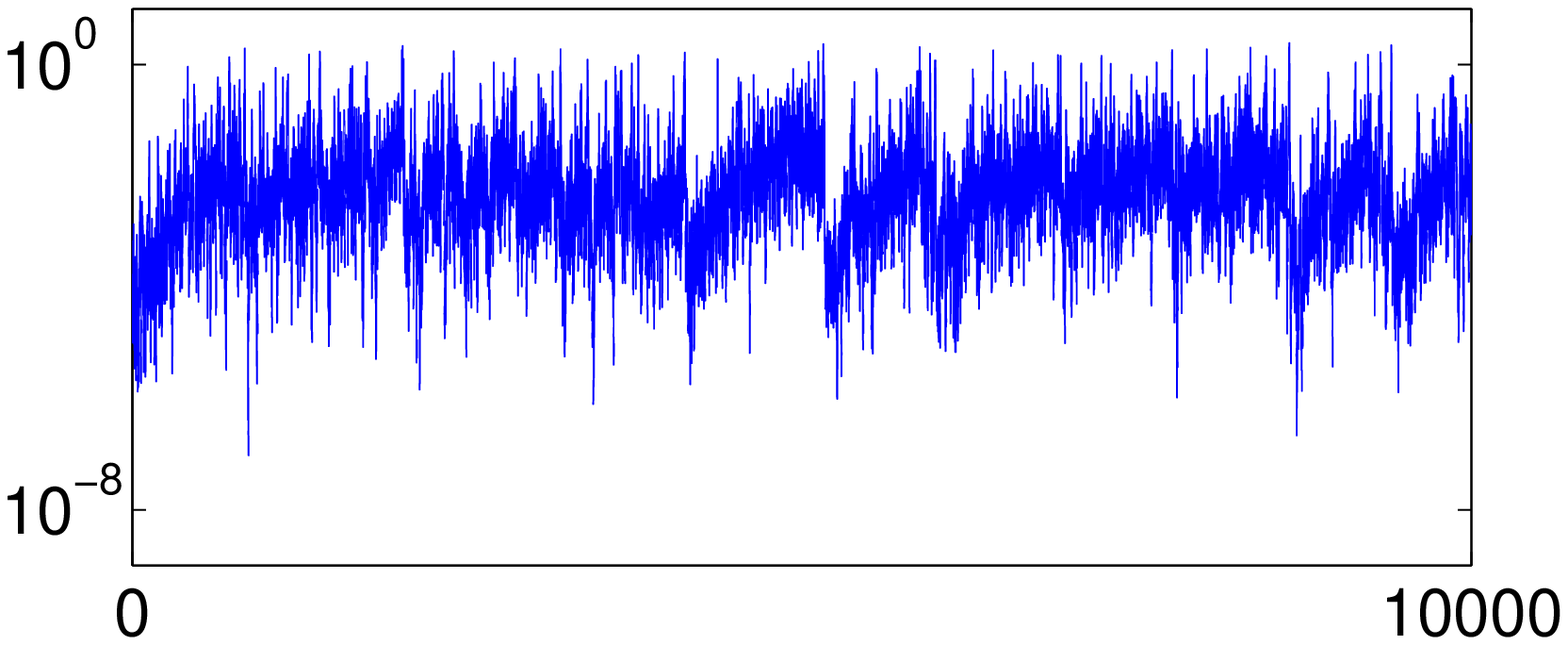}      }
\centerline{  (a) \hspace{5.5cm} (b) \hspace{5.5cm} (c)                }
\vspace{-0.5cm}
\end{center}
\caption{\label{fig1} Time series of the solution of eq. (\ref{model}) for $\mu=0 .01$. Top:  linear scale and bottom in log scale.
a) $F=0$ corresponding to a white noise; b) $F_{_{OU}}=\gamma Y^2/2$, Ornstein-Uhlenbeck noise with $\gamma=1.5$; c)  $F_{_{AN}}=\nu |Y|$, with $\nu=0.75$. Note the differences in the $y$-coordinate values.}
\end{figure*}

%

In the dynamo context, the turbulent fluctuations act as a multiplicative term in the equation for the magnetic field. In contrast to the case of equilibrium phase transition where additive thermal fluctuations prohibit phase transition in small dimensions, bifurcations are not destroyed by  multiplicative fluctuations even for  small (possibly zero) dimensions.  We thus start with a zero dimensional system subject to multiplicative noise.  For a multiplicative white noise, on-off intermittency is a generic behavior close to the threshold of instability \cite{on-off,on-off2}. Then the averaged amplitude  scales as $\langle X \rangle \propto \mu$. Although the exponent  differs from the mean-field prediction,  its value $\beta=1$ is in disagreement with the one measured in the experimental dynamo. 
It has been shown that  on-off behavior is observed when the departure from onset is smaller than half of the value of the noise spectrum at zero frequency \cite{sebonoff}.  
 In the dynamo experiment, on-off intermittency is not observed. We suggest that it is due to the absence of noise component at zero frequency and to strengthen this   hypothesis, we investigate the effect of a noise whose spectrum at zero frequency vanishes. We thus consider the dynamics of the unstable mode $X$ given by 
\begin{eqnarray}
\dot{X}&=&\mu X -X^{n+1}+\dot{Y} X\,,\nonumber\\
\dot{Y}&=&-F_Y+\zeta\,.
\label{model}
\end{eqnarray}

Here $\zeta$ is a Gaussian white noise with $\langle \zeta(t) \zeta(t') \rangle=2 \delta(t-t')$.
$F$ is a (potential) function of $Y$ and the subindex denotes differentiation with respect to this variable. 
$\dot{Y}$ acts as a multiplicative noise (for $X$) whose frequency-spectrum is controlled by the function $F(Y)$. 
When the potential $F$ is such that  the second moment of $Y$ is finite,  the spectrum of $\dot{Y}$  vanishes at low frequency (it behaves as the square of the frequency $f$, for small $f$). 
Standard estimates of the effect of noise on the onset of instability (for instance by calculating the evolution of the ensemble average of  $ \log X $ from the linear part of the first equation \cite{arnold}) show that the onset of instability of the solution $X=0$  is not affected by the noise and remains at $\mu=0$. In contrast, the non-linear regime above onset is strongly affected.
We display in fig. \ref{fig1} time series of $X$ for different functions $F$ in the vicinity of the onset of instability 
(unless otherwise stated, numerical simulations are performed in the case of cubic nonlinearities: $n=2$). 
For panel (a) we used  white noise, $F=0$ and on-off intermittency is observed: short bursts of finite amplitude (on-phases) alternate with long durations with negligible amplitude (off-phases).
%
In panel (b) the case $F=F_{_{OU}} \equiv \gamma Y^2/2$ is presented. For this choice $Y$ is the Ornstein-Uhlenbeck process. 
There is no off-phase and we expect a behavior for the moments that differs from the one of on-off intermittency. 
Panel (c) displays a time series for $F=F_{_{AN}}\equiv\nu |Y|$, that results in an intermediate behavior.

\begin{figure}[h!]
\begin{center}
\includegraphics[width=9cm,angle=0]{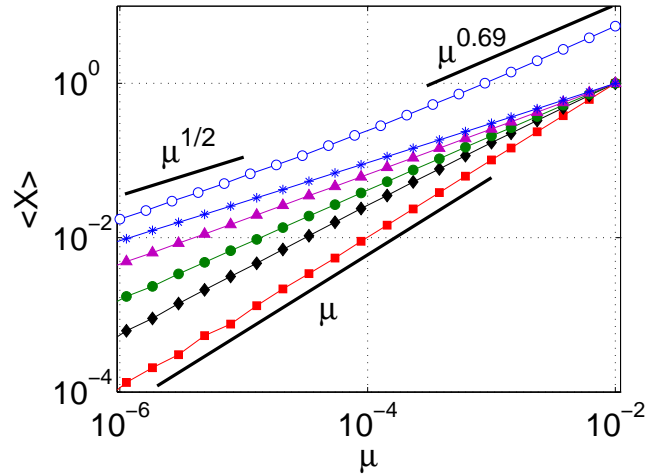}
\caption{\label{fig2} First moment $\langle X \rangle$ as a function of $\mu$ for  the solution of eq. (\ref{model}) for $F=F_{AN}$ with ($\blacksquare$) $\nu=0.125$, ($\blacklozenge$) $\nu=1.125$, ($\bullet$) $\nu=1.375$, ($\blacktriangle$) $\nu=1.75$, ($\bigstar$) $\nu=2.5$. The data are presented in loglog scale and have been normalized by their value at $\mu=0.01$.   The results for $F=F_{OU}$  with $\gamma=0.2$ ($\circ$) are presented and shifted for comparison. The  thick continuous lines indicate the exponents $1/2$, $0.69$ and $1$. }
\end{center}
\end{figure}

In fig. \ref{fig2}  the first moment is displayed as a function of $\mu$ for the two functions $F_{_{OU}}$ with $\gamma=0.2$ and 
$F_{_{AN}}$ with various values of $\nu$.
For the $F_{_{OU}}$ case we observe for $\mu \in [3.10^{-4},10^{-1}]$  an evolution that seems compatible with a  power law.  A best fit determination of the associated  exponent results in the value $0.69$,  thus different from $1$ and $1/2$.
However when $\mu$ is very small, the slope changes and the deterministic exponent $1/2$ is recovered: the apparent anomalous behavior 
disappears at criticality \cite{alexcomment}. This is confirmed by a perturbative expansion  performed on the Fokker-Planck equation (not presented here). This expansion predicts that $X$ is concentrated around the value $X^*$ at which a weighted average of the non-linear effect  
balances the linear growth rate $\mu = X^{*\,n} \int_{-\infty}^{\infty} \Pi(Y) \exp {(n Y)}\, dY$, where
$\Pi(Y)\propto e^{-F}$ is the stationary probability density of $Y$. 
Thus, in this case and for $n=2$, the first moment scales as $\sqrt{\mu}$ as  observed numerically.  


A simple potential $F$ for which this expansion can break down is $F_{_{AN}}=\nu|Y|$.
Indeed,  if $\nu>n$ the expansion holds resulting in normal scaling $\beta=1/n$  but breaks down (because $X^*$ vanishes) when $\nu<n$.
In fig. \ref{fig2}, where the first moment for this potential is displayed, 
we observe that $\langle X \rangle \propto \mu$  for small $\nu$ and $\langle X \rangle \propto \sqrt{\mu}$ 
for large $\nu$. Anomalous behavior with exponent between $1/2$ and $1$ is observed for 
$\nu$ of order $1$. In this regime and in contrast to the $F_{_{OU}}$ case, the exponent remains anomalous for the smallest achievable values 
of $\mu$. This numerical result is confirmed 
by a new perturbative expansion that we now sum up. 

Using $\Omega=\log X-Y-\log{\mu}/n$,  the Fokker-Planck equation for $P$ the stationary probability density function (p.d.f.)  of $\Omega$ and $Y$ is
\begin{equation}
0=-\mu \partial_{\Omega}(1-e^{n \Omega+n Y}) P+\partial_Y(F_Y P)+\partial_{Y}^2 P\,.
\label{FPeq}
\end{equation}
Since the derivative in $\Omega$ is multiplied by a small parameter (we are interested in the limit $\mu\rightarrow 0$), we introduce a WKB-like expansion and search for $P(\Omega,y) = \exp \left[ \sum_{m=-1}\mu^m S_m \right]$, where the first term $S_{-1}$ depends only on $\Omega$. 
At lowest order we obtain
\begin{equation}
\partial_Y^2 r_0+\partial_Y ( F_Y\, r_0)+ S_{-1,\Omega} (e^{n \Omega+n Y}-1) r_0=0\,,
\end{equation}
where $r_0=\exp [S_0]$. 
This equation can be solved exactly for positive and negative $Y$. 
The two solutions are then matched at $Y=0$ which selects the value of $S_{-1}$
\begin{equation}
n=2\nu I_\kappa \left[\lambda e^{n\Omega/2}\right] K_\kappa\left[ \lambda e^{n\Omega/2}\right]\,,
\label{lambdaeq}
\end{equation}
where $\lambda^2 = -4S_{-1,\Omega} /n^2$, $\kappa=\sqrt{\nu^2/n^2-\lambda^2}$
and $I_\kappa$ and $K_\kappa$ are modified Bessel functions of order $\kappa$.
The solution for $r_0$ is then

\begin{equation}
r_0 = \Biggl\{
\begin{array}{r}
A(\Omega) e^{-F/2}  I_\kappa\left[\lambda e^{n(\Omega +Y)/2}\right]\,K_\kappa\left[ \lambda e^{n \Omega/2} \right]\quad ({Y < 0}) \\ \, \\
A(\Omega) e^{-F/2}  K_\kappa\left[\lambda e^{n(\Omega +Y)/2}\right]\,I_\kappa\left[ \lambda e^{n \Omega/2} \right]\quad ({Y > 0}).
\end{array}
\end{equation}

\begin{figure}[h!]
\begin{center}
\includegraphics[width=8cm,angle=0]{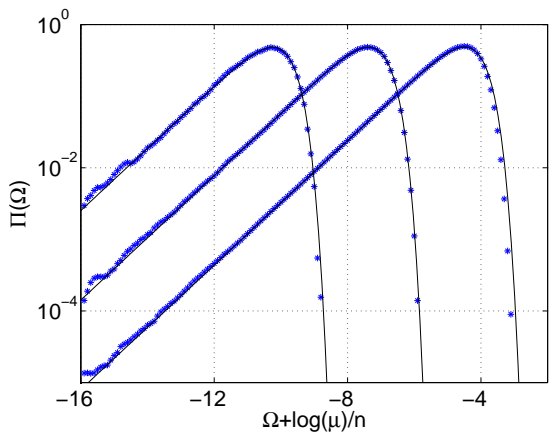}
\caption{\label{fig3} Probability density function $\Pi$ of $\Omega+\log(\mu)/n=\log(X)-Y$. The continuous line is the theoretical prediction and the symbols are the PDF calculated from the numerical solutions of the Langevin equation.  Here $\nu=1$ and the three curves are associated (from left to right) to $\mu=1.78 \, 10^{-5}$, $\mu=3.16\, 10^{-4}$ and $\mu=5.6 \, 10^{-3}$.}
\end{center}
\end{figure}

The amplitude $A(\Omega)$  is determined from the solvability condition at next order. Up to this order, we have then obtained the expression $P=\exp[\mu^{-1}S_{-1}(\Omega)] r_0(\Omega,Y)$ where all the dependence in $\mu$ is in the exponential.   As displayed in fig. \ref{fig3},  this asymptotic result  is in good agreement with the numerical simulations of the Langevin equations 
(\ref{model}).

From this formulation, we can calculate the moments. The exponential term acts as a cut-off for large $\Omega$ and is of the form $\exp{\left(-\mu^{-1}\exp{(n\,\nu \Omega/(n-\nu))} \right)}$. Therefore for $\mu\rightarrow 0$ and $\nu<n$ , only very negative $\Omega$ have to be considered. In this limit the amplitude $A(\Omega)K_\kappa\left[ \lambda e^{n \Omega/2}\right]$ tends to a constant and, after several standard estimates of the asymptotic behavior of the Bessel functions, we obtain for $\nu < n$  

\begin{equation}
\beta=\hbox{min} \left[\frac{1}{\nu},1 \right]\,.
\label{solexp}
\end{equation}

%
%
%
%

\begin{figure}[h!]
\begin{center}
\includegraphics[width=8cm,angle=0]{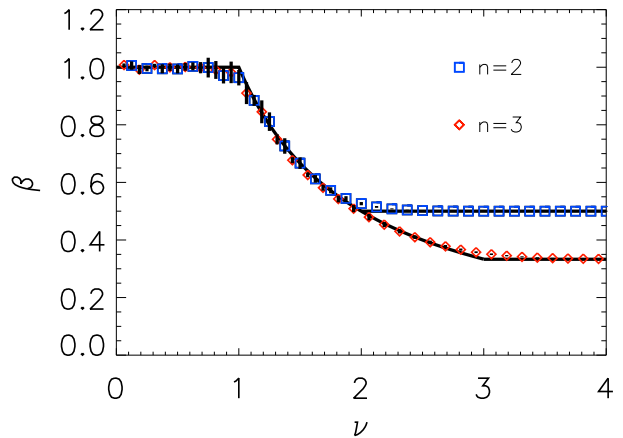}
\caption{\label{fig5} Exponents of the first moment as a function of  $\nu$ for ($\square$): $n=2$ and ($\diamond$): $n=3$. The continuous lines are the theoretical predictions and the symbols are obtained from the numerical solutions of the Langevin equation. }
\end{center}
\end{figure}

We tested our prediction by  numerically   calculating the first  moment
for different values of $\nu$ and for $n=2$ and  $n=3$. The results are shown in figure \ref{fig5}. For all cases the predictions are within
the error-bars of the numerically calculated values of $\beta$, and thus the predictions are verified.  To discuss one particular value,    the numerically computed exponent for $\nu=1.5$ and $n=2$ is $\beta=0.66\pm0.02$ which is in perfect agreement with the theoretical prediction $2/3$.  
We have also performed several numerical simulations using potentials of the form $F=-\nu \sqrt{Y_0^2+Y^2}$. We  have observed that only the behavior of
$F$ for large values of $|Y|$ is important.  In other words, the universality classes of the problem ({\it i.e.} the models having the same critical exponents) are determined by the behavior of the tails of 
$\Pi(Y)$. Incidentally, this shows that the anomalous scaling is not caused by the non-analyticity of $F$ at $Y=0$.

At this stage, we emphasize that our perturbative expansion (in $\mu$) allows to calculate an exact (non perturbative) expression for the value of the anomalous exponent. This exponent transitions from its on-off value $1$ for $\nu \le 1$ to its deterministic value $1/n$ for $\nu \ge n$. In the simple case of cubic nonlinearities, we predict an exponent between $1/2$ and $1$. Interestingly enough, the scaling reported in the dynamo experiment belongs to this range.

We have focused here on the first moment of the unstable mode. The behavior of higher moments is also of interest. It can be characterized by the set of exponents $\beta_p$ defined by $\langle X^p \rangle \propto \mu^{\beta_p}$. In the absence of fluctuations or at usual equilibrium phase transitions, monoscaling is  observed which means that $\beta_p=p\beta_1$.  The situation is richer here: there is no linear relation between the exponents (for instance it can be easily proved that $\beta_n=1$). Thus the solutions of model (\ref{model}) display multiscaling. This  is related to the complex structure of  the p.d.f. of $X$. In particular, it  cannot be expressed as a simple one-parameter distribution characterized by its first moment in contrast to the scaling hypothesis close to the critical point of an  equilibrium phase transition \cite{phasetrans}.

\begin{figure}[h!]
\begin{center}
\includegraphics[width=8.5cm,angle=0]{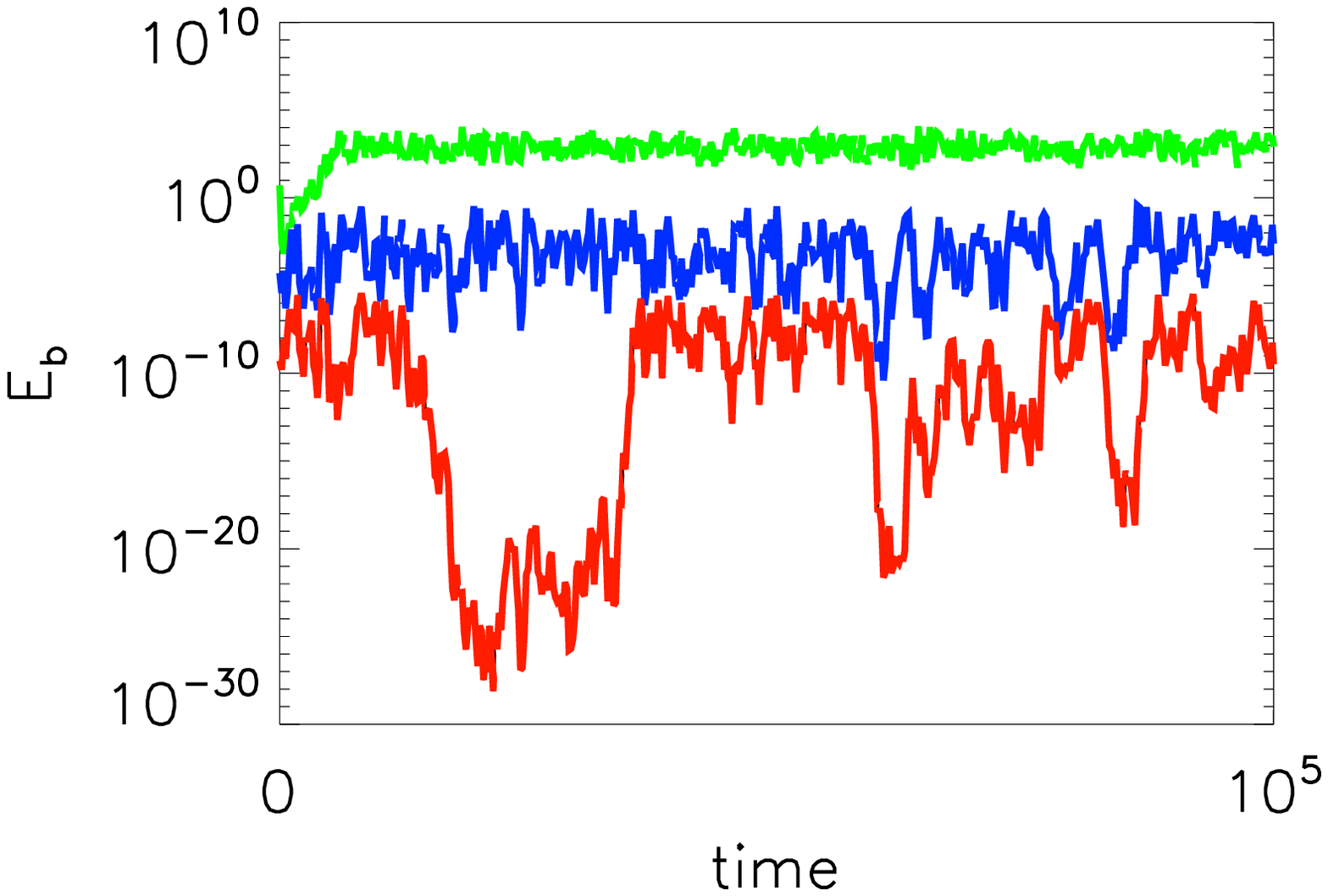}
\includegraphics[width=8cm,angle=0]{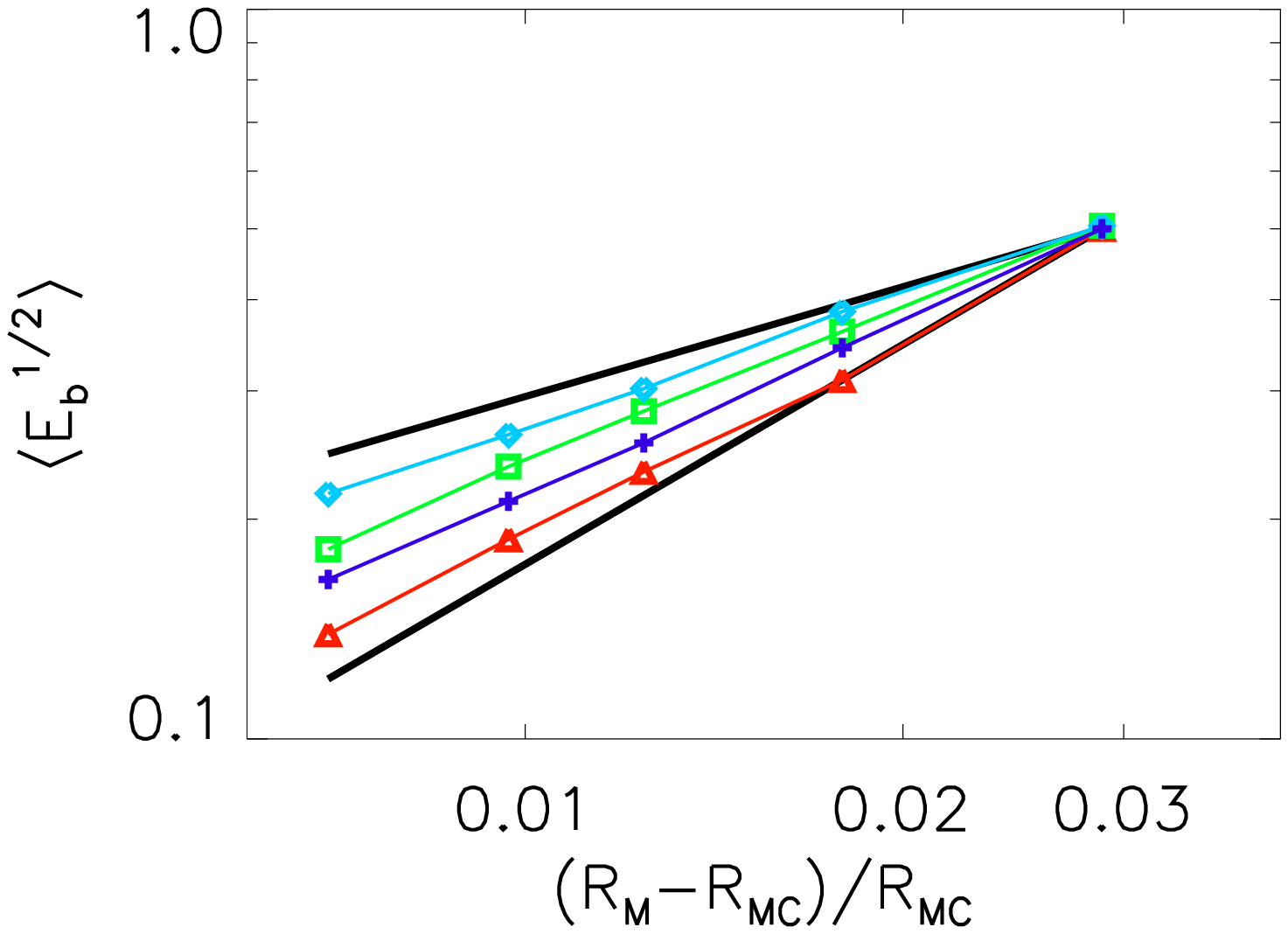}
\caption{\label{fig6} (Top) Time series of the space-averaged magnetic energy  $E_b=\bar{B^2}$ above the dynamo onset for (from top to bottom) $F=F_{OU}$ with  $\gamma=1$, $F=F_{AN}$ with $\nu=0.4$ and a white noise (see discussion in the text). The curves have been shifted for clarity. (Bottom) First moment as a function of the departure from onset $(Rm-Rm_c)/Rm_c$ for $F=F_{AN}$ with ($\triangle$): $\nu=0$,  ($+$): $\nu=0.1$,  ($\square$): $\nu=0.4$ and ($\diamond$):  $\nu=0.8$. The two thick lines indicate the exponents $1/2$ and $1$. }
\end{center}
\end{figure}

Another important issue is the effect of spatial dimension. The model
(\ref{model}) is zero dimensional ($X$ only depends on time and not on space) while the magnetic field in magnetohydrodynamics (MHD)
depends on three spatial dimensions. Analytical predictions for the
critical behavior at larger (non-zero) dimensions would be of great
interest but are still out of reach at present. To investigate
further the pertinence of our model to the dynamo instability, we have
performed direct numerical simulations of the MHD equations. 
To increase our control on the velocity temporal behavior, we used the
infinite Prandtl number limit \cite{alexPminfiny}. In this limit the
velocity is slaved to an external mechanical forcing and the Lorentz
force 
\[  \nabla^2 {\bf u} = {\bf F + b\cdot \nabla b - \nabla P}\,, \]
where ${\bf b}$ is the magnetic field and ${\bf F}$ is the body force. It  is proportional to the ABC flow 
 ${\bf F}= A_n[ 5\sin(z) + 2\cos(y),  2\sin(x) + 5\cos(z), 2\sin(y)+2\cos(x)] $ (see for instance \cite{ABC}). $A_n$ is an amplitude that  changes every time interval $\tau$ based on a
discrete version of our model $A_{n+1} =A_0 + (Y_{n+1}-Y_n)$ and $Y_{n+1}=Y_n -\tau F(Y_n) + r_n$ where $r_n$ is a random number. 
The magnetic field satisfies the induction equation
\begin{equation}
\frac{\partial {\bf b}}{\partial t}= \nabla \times \left( {\bf u \times b} \right)+Rm^{-1}\nabla^2 {\bf b}.
\nonumber
\end{equation}
The MHD equations were solved in a periodic box of size $2\pi L$ using a standard pseudo-spectral code \cite{gomez}  on a grid $32^3$. 
The magnetic Reynolds number defined by $R_m=\langle \|u\|^2\rangle^{1/2} L/\nu$ was varied above the onset value $Rm_c\simeq 11.65$.
In fig. \ref{fig6} (a), we display time series of the magnetic energy and note that they are similar to those presented in fig. \ref{fig1}. 
The first moments are displayed in fig. \ref{fig6} (b) for several values of $\nu$. We observe that the exponent of the first moment decreases
from $1$ to $1/2$ when $\nu$ increases. Estimates of the exponent are computationally demanding so that a quantitative comparison with  
our model is out of reach. Nevertheless, the results reported here support the robustness of the behavior we have identified.

To summarize, we have presented a simple model that results in anomalous exponents which lie between the 
deterministic value and the on-off intermittent one. The exact value of these exponents was calculated using an asymptotic expansion.   
The model emphasizes the role of the noise spectrum at zero frequency. It remains to be understood whether and when turbulent fluctuations can be modeled as the noise considered here \cite{scal}.
In addition,  how such a noise affects other phase transitions and whether the present expansion can capture other critical exponents are interesting  open questions.

We greatly acknowledge Stephan Fauve for rising our interest on this topic \cite{GAFD} and also for several discussions and constant support. 
Computations were carried out on the CEMAG computing center at LRA/ENS and on the CINES computing center, and 
their support is greatly acknowledged.


\begin{thebibliography}{5}
\bibitem{alexcomment} The crossover between normal and apparently anomalous scaling occurs for a value $\mu_{c}$ that tends to zero when $\gamma$ vanishes. Thus the dependence of $\beta$ on $\gamma$ is discontinuous at  $\gamma=0$.
\end{thebibliography}
\end{document}